\documentclass[
aps,%
10pt,%
final,%
notitlepage,%
oneside,%
twocolumn,%
nobibnotes,%
nofootinbib,%
superscriptaddress,%
noshowpacs,%
centertags]%
{revtex4}

\usepackage[cp1251]{inputenc}
\usepackage{amsmath}
\usepackage{color}
\usepackage{amsfonts}
\usepackage{epsfig}
\usepackage{graphics}

\begin{document}

\title{Scalar field with the source in the form of the stress-energy tensor trace as a dark energy model}

\author{\firstname{I. G.}~\surname{Dudko}}

\email{igordudko89@gmail.com} \affiliation{B. I. Stepanov
Institute of Physics, Belarus }
\author{\firstname{Yu. P.}~\surname{Vyblyi}}
\email{vyblyi@gmail.com} \affiliation{B. I. Stepanov Institute of
Physics, Belarus}
%\date{\today}
%\today
\begin{abstract}
We consider a scalar-tensor theory of gravitation with the scalar
source being the trace of the stress-energy tensor of the scalar
field itself and matter. We obtain an example of a numerical
solution of the cosmological equations which shows that under some
special choice of the scalar parameters, there exists a slow-roll
regime in which the modern values of the Hubble and deceleration
parameters may be obtained.
\end{abstract}

\maketitle

\section{Introduction}
As well known, that the accelerated expansion of the Universe is
successfully described by the $\Lambda CDM$ model. However, the
nature of the $\Lambda$-term is unclear and the physical
interpretation of this constant led us to the famous problem of
$120$ orders \cite{{Sahni},{Padma}}. For this reason, in the
literature are discussed various alternative models of the dark
energy. Scalar fields are often considered as candidates for the
dark energy (see, for example, \cite{{Matos2},{Matos1},{Peeble},
{Copeland}} or brief but enough exhausting review in \cite{Bel}).
Nevertheless, there is no unambiguous criterion for the choice of
the field Lagrangian in the scalar field theories. Moreover, as it
was shown in \cite{Scherrer}, any scalar field in the slow-roll
regime can simulate the cosmological constant and hence lead to an
appropriate cosmological scenario.

 As known, the Einstein equations of the gravitational field can be
derived if one considers a tensor field in Minkowski space-time
with the source being the total stress-energy tensor of both field
and matter  \cite{{Deser},{Wei}}. In the present work, we consider
a scalar field with a similar property: assuming that the scalar
field source is the trace of the stress-energy tensor of both
matter and the field itself and that this scalar field minimally
interacts with gravitational field. In the frameworks of this
approach derive receiving the scalar field Lagrangian which
contains three parameters: new constant of the scalar interaction,
scalar field mass and a parameter which relates to the minimum  of
scalar potential \cite{Vyb}. Note that the task to get such free
scalar Lagrangian was considered many years ago by P. Freund and
Y.Nambu in \cite{Fre}. Our Lagrangian generalizes that one from
\cite{Fre} and besides we consider the interaction of the scalar
field with the matter.

\section{Scalar field equation}
In this article, we use the gravitational system of units $G=c=1$.
Let us consider a scalar field $\varphi$ with the source being the
total trace of the stress-energy tensor of the scalar field and
matter fields. This condition implies that the scalar field
equation has the following form
\begin{equation}\label{fieldeq0}
(\square-m^2)\varphi=q (T^{\varphi}+T^{M}),
\end{equation}
the d'Alembertian is
$\square=-\nabla_{\mu}\nabla_{\nu}g^{\mu\nu}$, the constants $m$
and $q$ mean the mass and the interaction constant of the scalar
field\footnote{The metric signature $(+,-,-,-)$ is used}.

Equation (\ref{fieldeq0}) allows to determine the Lagrangian of
the scalar field and matter \cite{Vyb}
\begin{eqnarray}\label{lag}
L_{\varphi}+L_M=\frac{1}{2}(\frac{\partial_{\mu}\varphi\partial^{\mu}\varphi}{1+2q\varphi}
-\nonumber\\
\left.-m^2\varphi^2+C\left(1+2q\varphi\right)^2\right)\sqrt{-g}+\nonumber\\
+L_M(\left(1+2q\varphi\right)g_{\mu\nu},Q_M),
\end{eqnarray}
where $Q_M$ stands for matter fields. Note that in \cite{Fre} the
term with constant $C$ was absent. The total Lagrangian of the
model with minimal interaction between scalar and gravitational
field takes the form

\begin{equation}
L=L_g+L_{\varphi}+L_M,
\end{equation}
where
\begin{equation}
L_g=R\sqrt{-g}
\end{equation}
is the Lagrangian of gravitational field.

Lagrangian (\ref{lag}) has non-linear kinetic energy term an
belong to so-called k-essence model \cite{Arm-Pic}.

 The scalar field equation
and his stress-energy tensor have the form
\begin{eqnarray}\label{fieldeq3}
\varphi^{;\mu}_{;\mu}+m^2\varphi=q(\frac{\varphi_{,\mu}\varphi^{,\mu}}{1+2q\varphi}-
2m^2\varphi^2+\nonumber\\+2C(1+2q\varphi)^2-T^{M}),
\end{eqnarray}
\begin{eqnarray}\label{T}
T^{\mu\nu}_{\varphi}=\frac{\partial^{\mu}\varphi\partial^{\nu}\varphi}{\Phi}-\frac{1}{2}g^{\mu\nu}(\frac{\partial_{\alpha}\varphi
\partial^{\alpha}\varphi}{\Phi}-\nonumber\\
\left.-m^2\varphi^2+C\Phi^2\right).
\end{eqnarray}

So, the interaction between the matter and the scalar field is
realized with the help of the effective metric
\begin{equation}
f_{\mu\nu}=(1+2q\varphi)g_{\mu\nu}=\Phi g_{\mu\nu}.
\end{equation}

The scalar field, interacting with the matter, was investigated in
several papers (see, for example, \cite{{Sot}, {Don}}), in
particular within the framework of chameleon model \cite{{Khoury},
{Das}}.

Let us consider the scalar field in the state with the minimal
energy. The minimum of the energy is achieved at the point where
the potential
\begin{equation}\label{potential0}
V(\varphi)=\frac{1}{2}m^2\varphi^2-\frac{C}{2}(1+2q\varphi)^2
\end{equation}
has a minimum. Here $\varphi$ is assumed to vary from $-1/2q$ to
$+\infty$, which ensures the positivity of the denominator in
Lagrangian (\ref{lag}). Under the condition $C<m^2/4q^2$,
potential has the minimum at the point
\begin{equation}
\varphi_0=\frac{2qC}{m^2-4q^2C},
\end{equation}
\begin{equation}
V(\varphi_0)=-\frac{m^2C}{2(m^2-4q^2C)}.
\end{equation}
For $C<0$, the minimum value of the potential is positive:
$V(\varphi_0)>0$, therefore it can be identified with the
cosmological constant in the Einstein equations.

For $0<C<m^2/4q^2$, the minimum value of the potential is
negative: $V(\varphi_0)<0$. In this case, $|V(\varphi_0)|$ can be
interpreted as the squared graviton mass $\mu^2$, since the
Einstein equations in the linear approximation can be written as
[11]
\begin{equation}
\left(\square+V(\varphi_0)\right)\psi_{\mu\nu}=V(\varphi_0)\gamma_{\mu\nu}.
\end{equation}

Here, in contrast to the free massive Fierz-Pauli equation in the
Minkowski space, the term $V(\varphi_0)\gamma_{\mu\nu}$ is the
stress-energy tensor of the scalar field in the ground state.

Assuming that $|C|$ is small as compared with $m^2/q^2$ we get the
estimation $\mu^2\approx |C|/2$.

\section{Cosmological solution}

Let us consider the application of the scalar-tensor theory of
gravity to cosmology. As a first approximation we consider the
field equations without interaction of the scalar field with
matter. Thus the Lagrangian of the matter takes the form
\begin{equation}
L_M\left(f_{\mu\nu},Q_M\right)\rightarrow
L_M\left(g_{\mu\nu},Q_M\right).
\end{equation}

The scalar factor vanishes from stress-energy tensor of the matter
in the equations of gravitational, field and stress-energy tensor
trace vanishes from scalar field equation. Thus system of
equations takes the form
\begin{eqnarray}
G^{\mu\nu}=8\pi(\frac{\partial^{\mu}\varphi\partial^{\nu}\varphi}{\Phi}-\frac{1}{2}\frac{\partial_{\alpha}\varphi
\partial^{\alpha}\varphi}{\Phi}g^{\mu\nu}+\nonumber\\
+\frac{1}{2}m^2\varphi^2g^{\mu\nu}
-\frac{1}{2}C\Phi^2g^{\mu\nu}+\nonumber\\
\left.+\left(\epsilon+p\right)u^{\mu}u^{\nu}-pg^{\mu\nu}\right),
\end{eqnarray}
\begin{eqnarray}
\left(\Box-m^2\right)\varphi=q(-\frac{\partial_{\mu}\varphi\partial^{\mu}\varphi}{\Phi}+\nonumber\\
\left.+2m^2\varphi^2-2C\Phi^2\right),
\end{eqnarray}

Spatially flat homogeneous and isotropic universe is described by
Friedman-Lemaitre-Robertson-Walker the metric
\begin{eqnarray}
ds^2=dt^2-a^2\left(t\right)\left(dr^2+r^2d\theta^2+\right.\nonumber\\
\left.+r^2\sin^2\theta d\phi^2\right).
\end{eqnarray}
Cosmological equations, which include Friedman equation, massless
scalar field equation and covariant law of conservation of
stress-energy tensor of matter, read
\begin{eqnarray}\label{cos}
\left(\frac{\dot{a}}{a}\right)^2=\frac{8\pi}{3}\left(\frac{1}{2}\frac{\dot{\varphi}^2}{\Phi}+\frac{1}{2}m^2\varphi^2-\right.\nonumber\\
\left.-\frac{1}{2}C\Phi^2+\epsilon\right),
\end{eqnarray}
\begin{equation}\label{cos1}
\ddot{\varphi}+3H\dot{\varphi}-\frac{q\dot{\varphi}^2}{\Phi}+m^2\varphi\Phi-2Cq\Phi^2=0,
\end{equation}
\begin{equation}\label{cos2}
\dot{\epsilon}+3H\left(\epsilon+p\right)=0.
\end{equation}
Here $H\equiv \dot{a}/a$ is the Hubble parameter.

Let us consider the system of equations with the massless scalar
field (\ref{cos}--\ref{cos2}). Comparing the stress-energy tensor
of the scalar field with stress-energy tensor of the ideal fluid,
we can write the following relations
\begin{equation}
\epsilon^{\varphi}=\frac{1}{2}\left(\frac{\dot{\varphi}^2}{\Phi}-C\Phi^2\right),
\end{equation}
\begin{equation}
p^{\varphi}=\frac{1}{2}\left(\frac{\dot{\varphi}^2}{\Phi}+C\Phi^2\right).
\end{equation}
If we identify the density of the dark energy at present stage
with the energy density of the scalar field and specify parameter
$\omega$ in the equation of state we can get the initial values
$\varphi_0$ and $\dot{\varphi}_0$ for the equation of the scalar
field
\begin{equation}\label{nach}
\epsilon^{\varphi}_0\left(1+\omega\right)\equiv
a=\frac{\dot{\varphi}_0^2}{\Phi_0},
\end{equation}
\begin{equation}\label{nach1}
\epsilon^{\varphi}_0\left(1-\omega\right)\equiv b=-C\Phi_0^2,
\end{equation}
where $\Phi_0=1+2q\varphi_0$. Further, we will omit the index $0$.
One can consider two cases, which correspond to different
parameters $\omega$ -- $\omega>-1$ and $\omega<-1$. Then from the
system of equations (\ref{nach}), (\ref{nach1}) we obtain two
solutions
\begin{equation}\label{Q}
\omega>-1,\quad a>0,\quad \varphi_q\geq-\frac{1}{2q}
\end{equation}
and
\begin{equation}\label{ph}
\omega<-1,\quad a<0,\quad \varphi_{ph}\leq-\frac{1}{2q}
\end{equation}
This case corresponds to so-called phantom dark energy
\cite{{Caldwell},{Caldwell et al.}}.

 From equation (\ref{nach1}) we get
\begin{equation}
\varphi_{1,2}=-\frac{1}{2q}\pm\frac{1}{2q}\sqrt{\frac{b}{C'}},
\end{equation}
which implies that $C<0$, and $C'\equiv -C>0$. For $\dot{\varphi}$
we have
\begin{equation}\label{speed}
\dot{\varphi}=\pm\left(\pm
a\sqrt{\frac{b}{C'}}\right)^{\frac{1}{2}}.
\end{equation}
Taking into account the inequalities (\ref{Q}) and (\ref{ph}), we
obtain the following initial values $\varphi$ and $\dot{\varphi}$
for the $\omega>-1$:
\begin{equation}
\varphi_q=-\frac{1}{2q}+\frac{1}{2q}\sqrt{\frac{\epsilon\left(1-\omega\right)}{C'}},
\end{equation}
\begin{equation}
\dot{\varphi}_q=\left(\epsilon\left(1+\omega\right)\sqrt{\frac{\epsilon\left(1-\omega\right)}{C'}}\right)^{\frac{1}{2}},
\end{equation}
and for the $\omega<-1$:
\begin{equation}
\varphi_{ph}=-\frac{1}{2q}-\frac{1}{2q}\sqrt{\frac{\epsilon\left(1-\omega\right)}{C'}},
\end{equation}
\begin{equation}
\dot{\varphi}_{ph}=\left(-\epsilon\left(1+\omega\right)\sqrt{\frac{\epsilon\left(1-\omega\right)}{C'}}\right)^{\frac{1}{2}}.
\end{equation}

\section{Cosmological scenario investigation}

At present stage, a scalar field must be in the slow roll mode
when the following relation is satisfied:
\begin{eqnarray}\label{slow}
3H\dot{\varphi}=2Cq\Phi= V'\Phi,
\end{eqnarray}
where $V'\equiv\frac{\partial V}{\partial\varphi}$. Expression
(\ref{slow}) was obtained under the assumption that
$q\dot{\varphi}^2/\left(1+2q\varphi\right)$ is small. This term
can be neglected in comparison with $3H\dot{\varphi}$ because of
smallness of $q$ and $\dot{\varphi}^2$. This slow roll mode can be
used for establishing relation between constants $C$ and $q$. One
can substitute the initial values of $\varphi$ and $\dot{\varphi}$
in (\ref{slow}). Then, using the dimensionless quantities
$\overline{C}$ and $\overline{\epsilon_{\varphi}}$, we obtain an
expression binding the constants $\overline{C}$ and $q$
\begin{equation}\label{C}
\overline{C'}=\frac{C'}{H^2}=\frac{81\left(1+\omega\right)^2}{16q^4\left(1-\omega\right)^3\overline{\epsilon}_{\varphi}},
\end{equation}
where $\overline{\epsilon}_{\varphi}\equiv\epsilon_{\varphi}/H^2$
 dimensionless energy density of the scalar field. The relation
(\ref{slow}) allows us to establish the sign of the initial value
$\dot{\varphi}$, namely $\dot{\varphi}<0$. Thus with an
appropriate choice of the constant $q$, the slow-rolling is
provided automatically both in two models of the dark energy. The
rolling speed $\dot{\varphi}$ depends only on $q$, as follows from
(\ref{speed}) and (\ref{C}).

For the numerical solving of system of equations
(\ref{cos}--\ref{cos2}) we have to pass to dimensionless
quantities in this system. Let us define the dimensionless time as
$T=tH_0$. Hence $T=1$ means a period of time equal to one age of
the Universe. Then the system of cosmological equations takes the
form
\begin{eqnarray}
\left(\frac{a'}{a}\right)^2=\frac{8\pi}{3}\left(\frac{(\varphi')^2}{2\Phi}-\overline{C}\Phi^2+\overline{\epsilon}\right),
\end{eqnarray}
\begin{eqnarray}
\varphi''+3\overline{H}\varphi'-q\frac{(\varphi')^2}{\Phi}+\nonumber\\
+2\overline{C}q\Phi^2=0,
\end{eqnarray}
\begin{equation}
\epsilon'+3\overline{H}\left(\epsilon+p\right)=0.
\end{equation}
We have one arbitrary parameter $q$. The numerical solution of the
cosmological equations we will obtain when the value of the
parameter equals $q=1$. In SI units this value is several orders
of magnitude less than the gravitational constant which ensures
the absence of symptoms of a scalar field in the solar system. The
numerical solution of these equations is shown in figures.

\begin{figure}[h]
\centering
\includegraphics[scale=0.8]{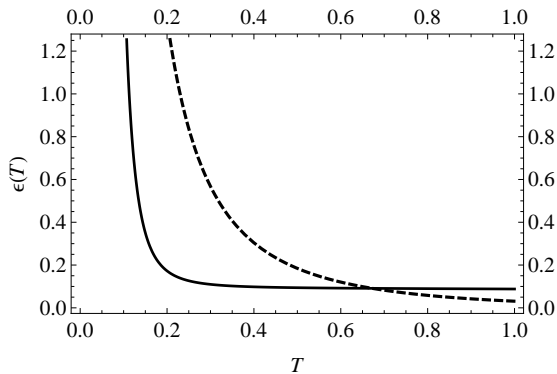}
\caption{Energy density of scalar field $\epsilon_{\varphi}(t)$
and matter $\epsilon_M(t)$ dependence on time. Continuous line
corresponds to energy density of scalar field
$\epsilon_{\varphi}(t)$ and dashed line corresponds to energy
density of matter $\epsilon_M(t)$. The point $T=0$ corresponds to
the initial singularity and the point $T=1$ corresponds to the
present time for all dependence on time plots.}
\end{figure}

The figure $1$ shows decreasing the energy density of scalar field
and matter with time. Radiation energy density can be neglected
because its contribution to the total energy density of the
universe is very small in the considered period of time. Moment of
nearly equal contribution of energy densities of scalar field and
matter corresponds to dimensionless time $T=0.672$ and red-shift
$z=0.431$. The same values for $\Lambda$CDM model are $T=0.68$ and
$z=0.417$.

\begin{figure}[h]
\centering
\includegraphics[scale=0.8]{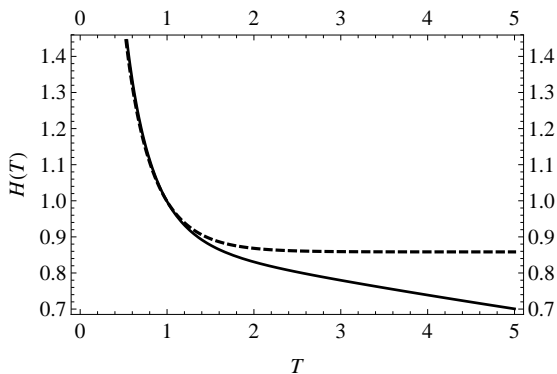}
\caption{Hubble parameter $H(t)$ dependence on time. Continuous
line corresponds to Hubble parameter which follows from our model
and dashed line corresponds to $\Lambda$CDM Hubble parameter.}
\end{figure}

\begin{figure}[h]
\centering
\includegraphics[scale=0.8]{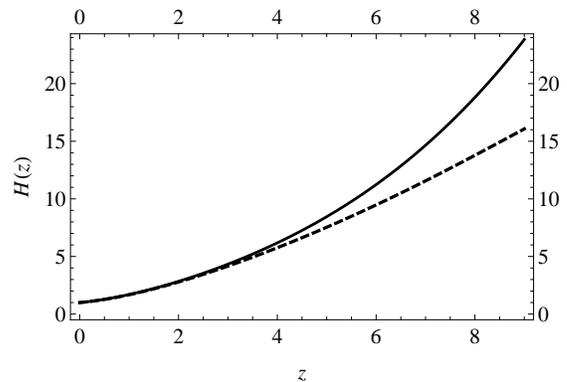}
\caption{Hubble parameter $H(z)$ dependence on red-shift.
Continuous line corresponds to Hubble parameter which follows from
our model and dashed line corresponds to $\Lambda$CDM Hubble
parameter.}
\end{figure}

From the figures $2$ and $3$ we see that the value of the Hubble
parameter at the present time which follows from our model
coincides with the observed value of Hubble constant and value
which follows from $\Lambda$CDM model. In dimensionless units
these values corresponds to $(\overline{H}(1)=H(1)/H_0=1)$.
However previous and further evolution of Hubble parameter in
different models is different. From $H(z)$ dependence one can see
that our model gives much bigger Hubble parameter in the past than
$\Lambda$CDM model. On the other hand $H(t)$ dependence shows us
that Hubble parameter of $\Lambda$CDM model will be constant when
the contribution of energy density of matter can be neglected.
This state of Universe will be corresponds to de-Sitter Universe
in which the energy density of $\Lambda$-term only presents . Our
model, on the contrary, gives a different result -- Hubble
parameter decreases with time.
\begin{figure}[h]
\centering
\includegraphics[scale=0.8]{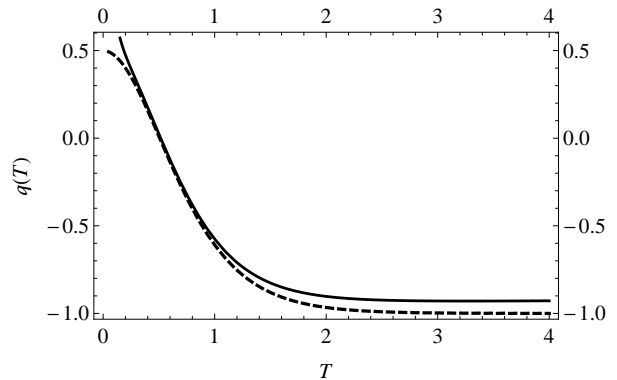}
\caption{The deceleration parameter $q(t)$ dependence on time.
Continuous line corresponds to deceleration parameter which
follows from our model and dashed line corresponds to $\Lambda$CDM
deceleration parameter.}
\end{figure}

\begin{figure}[h]
\centering
\includegraphics[scale=0.8]{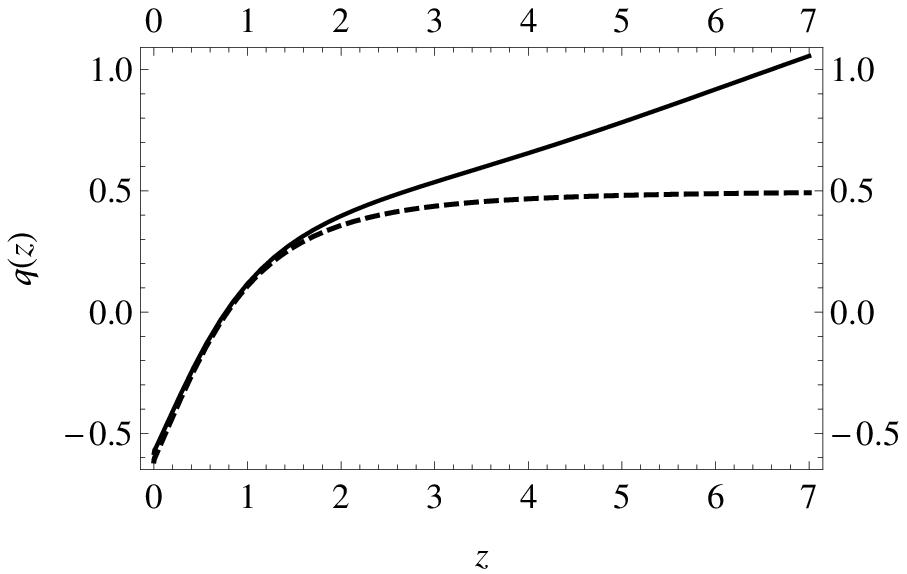}
\caption{The deceleration parameter $q(z)$ dependence on
red-shift. Continuous line corresponds to deceleration parameter
which follows from our model and dashed line corresponds to
$\Lambda$CDM deceleration parameter.}
\end{figure}

The figure 3 shows that the present value of deceleration
parameter $q(t)$ which follows from our model is in good agreement
with the value, which follows from $\Lambda$CDM model. Our model
gives $q=-0.574$ and $\Lambda$CDM model gives $q=-0.61$ for
$\Omega_M=0.26$ and $\Omega_{\Lambda}=0.74$. The moment when the
Universe stopped slowing down and began to accelerate $q=0$
corresponds to $z=0.767$ for our model and $z=0.785$ for
$\Lambda$CDM model. When $q$ is selected the parameter
$\omega_{\varphi}$ can be obtained from expression (\ref{T}). It
is important to note that parameter $\omega_{\varphi}$ grows and
begins positive, that leads to deceleration of cosmological
expansion.

\begin{figure}[h]
\centering
\includegraphics[scale=0.8]{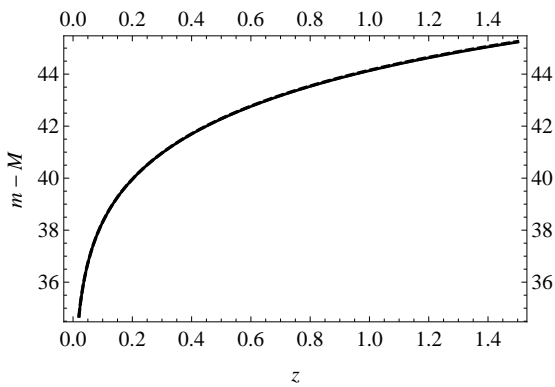}
\caption{The luminosity distance $D_{lum}(z)$ dependence on
red-shift. Continuous line corresponds to the luminosity distance
which follows from our model and dashed line corresponds to
$\Lambda$CDM the luminosity distance. This plot shows the
luminosity distance curves from $z=0$ to $z=1.5$ red-shift scale.}
\end{figure}

\begin{figure}[h]
\centering
\includegraphics[scale=0.8]{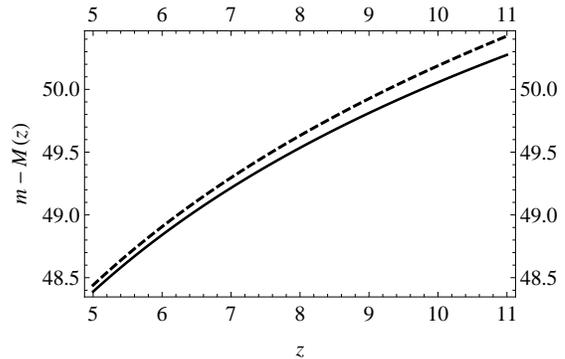}
\caption{The luminosity distance $D_{lum}(z)$ dependence on
red-shift. Continuous line corresponds to the luminosity distance
which follows from our model and dashed line corresponds to
$\Lambda$CDM the luminosity distance. This plot shows the
luminosity distance curves from $z=5$ to $z=11$ red-shift scale.}
\end{figure}

The figure $6$ shows that the plots of both model overlapping in
the red-shift scale from $z=0$ to $z=1.5$. It means that our model
in good agreement with the supernova observational data like
$\Lambda$CDM model. On the other hand figure $7$ of large
red-shift scale shows the differences between the models. In our
model the supernova are brighter than in the $\Lambda$CDM model at
the same large red-shift. This means that in our model supernova
are closer than in the $\Lambda$CDM model.

\section{Conclusion}
In  present work the nonlinear scalar field interacting with the
gravitational field and matter is introduced.  The requirement
that the source of the field is the trace of its own stress-energy
tensor leads to the Lagrangian containing three arbitrary
parameters. An example of numerical solution of the full system of
field equations, describing the cosmological solution for the
homogenous and isotropic Universe, shows that for certain
restrictions on the parameters the slow-roll regime is possible.
In this regime the scalar field simulates the dark energy in
agreement with the modern observational data. More detailed
analysis allowing for scalar field mass and interaction with the
matter for  different stages of cosmological expansion will be
carry out in future. \noindent

\newpage
%\small

\end{document}